\documentclass[journal=nalefd,manuscript=letter]{achemso}

\usepackage[version=3]{mhchem} 

\author{Axel J. M. Deenen}
\author{Dirk Grundler}
\affiliation{Laboratory of Nanoscale Magnetic Materials and Magnonics, Institute of Materials (IMX), School of Engineering, École Polytechnique Fédérale de Lausanne (EPFL), Lausanne, 1015, Vaud, Switzerland}
\altaffiliation{Institute of Electrical and Micro Engineering, School of Engineering, École Polytechnique Fédérale de Lausanne (EPFL), Lausanne, 1015, Vaud, Switzerland}
\email{Dirk.Grundler@epfl.ch}

\title[An \textsf{achemso} demo]
  {Superconducting Diode Effect due to Chiral Meissner Currents in a Hollow Superconducting Helix}

\abbreviations{IR,NMR,UV}

\begin{document}



\begin{abstract}
The superconducting diode effect (SDE) is a key nonreciprocal phenomenon with broad relevance for superconducting electronics. Using time-dependent Ginzburg-Landau simulations, we predict and quantify a superconducting diode effect arising solely from geometric chirality imposed to a conventional superconductor. The helical geometry and magnetic-field-induced screening currents produce inequivalent critical currents for opposite polarities. The diode efficiency reaches a maximum when one current direction first nucleates vortices, revealing a chirality-controlled crossover between screening- and vortex-dominated nonreciprocity. These results establish mesoscopic geometric chirality as a robust mechanism for supercurrent rectification in an achiral superconductor. They suggest an experimentally accessible route towards 3D superconducting diodes for multi-level integrated quantum circuits.
\end{abstract}


\section{Introduction}

The superconducting diode effect (SDE) describes unequal critical currents for opposite current directions. This non-reciprocity has garnered increased attention recently \cite{ando_observation_2020, nadeem_superconducting_2023} as the superconducting diode holds a similar role as the semiconductor diode has in current microelectronics\cite{ingla-aynes_efficient_2025, castellani_superconducting_2025}. The origin of the SDE can be multifold, ranging from finite momentum Cooper pairs\cite{davydova_universal_2022}, noncentrosymmetric crystal structures\cite{wakatsuki_nonreciprocal_2017}, to inhomogeneous magnetic fields \cite{edwards_superconducting_1962, swartz_asymmetries_1967} and systems with asymmetric vortex barriers \cite{vodolazov_superconducting_2005, carapella_bistable_2009, gutfreund_direct_2023, hou_ubiquitous_2023, moll_geometrical_2025}. Basic symmetry considerations state that when spatial-inversion and time-reversal symmetries are broken, transport possesses nonreciprocal characteristics \cite{rikken_observation_1997, rikken_electrical_2001, tokura_nonreciprocal_2018}. Spatial symmetry breaking is typically achieved by heterostructure engineering or the introduction of spin-orbit coupling (SOC), whereas time-reversal is broken by a magnetic field. In a chiral system, which inherently lacks spatial inversion symmetry, the magnetic field gives rise to electrical magnetochiral anisotropy (eMChA)\cite{rikken_observation_1997, rikken_electrical_2001}. It predicts that the resistance $R$  for right (D) or left (L) handed systems has a component that scales as $R^{D/L}(I)\propto \gamma^{D/L}\mathbf{B}\cdot\mathbf{I}$, with $\gamma^D=-\gamma^L$ the magnetochiral coefficient, $\mathbf{B}$ the magnetic field, and $\mathbf{I}$ the current. Consequently, the resistance is nonreciprocal: $R^{D/L}(I)=-R^{D/L}(-I)$ and $R^{D}(I)=-R^{L}(I)$. When applied to superconductors close to $T_{\rm c}$, eMChA predicts inequivalent critical currents\cite{nadeem_superconducting_2023}.  It was shown experimentally\cite{qin_superconductivity_2017} and later proven theoretically using time-dependent Ginzburg-Landau formalism \cite{he_supercurrent_2023, li_microscopic_2025} that a superconducting transition metal dichalcogenide chiral nanotube in a magnetic field exhibited a finite SDE, without needing SOC. The effect originated from the interplay between magnetic flux induced screening currents and the transport current. The diode efficiency was a periodic function of applied flux, similar to the Little-Parks effect\cite{little_observation_1962}, and on the order of a few percent close to $T_{\rm c}$. Subsequent works indicated an enhanced diode effect near 20\% by lowering the temperature, changing the radius, and introducing strain \cite{li_microscopic_2025}. However, so far, the  transition metal dichalcogenide chiral nanotube has been the only chiral structure for which a diode effect without SOC was demonstrated \cite{cordoba_topological_2024}.
Meanwhile, there have been major advancements in the 3D nanofabrication of superconductors\cite{makarov_new_2022} enabling curved surfaces with feature sizes down to below intrinsic length scales of superconductors\cite{orus_superconducting_2022}. An asymmetrically structured superconductor has recently shown a field-tunable vortex diode behavior\cite{moll_geometrical_2025}, and holds great potential for integration in future 3D multilevel interconnected circuits \cite{ingla-aynes_efficient_2025, castellani_superconducting_2025}. However, the interplay between artificially engineered chirality \cite{MXu2025} and superconductivity in curved 3D nanostructures \cite{cordoba_topological_2024} remains largely unexplored, both experimentally and theoretically. In particular, it is unknown whether purely geometric chirality, without SOC or intrinsic crystal asymmetry, generates a sizeable SDE in the experimentally accessible mesoscopic superconductors.

Stimulated by the recent advancements in nanofabrication, we explore the transport properties of a mesoscopic superconducting helix and bridge the gap between the SDE in the crystalline chiral nanotubes and the planar superconductor diodes. The helix configuration naturally combines geometric chirality with tunable curvature and torsion. It therefore provides an ideal platform to investigate the SDE in engineered chiral structures, an open subject which has not yet been addressed. We show that hollow mesoscopic type-II superconducting helices display a field- and chirality-dependent SDE. The origin of this lies in magnetic-field-induced chiral screening currents superimposed to directional transport currents. Our results extend beyond previous chiral-nanotube  \cite{qin_superconductivity_2017, he_supercurrent_2023} and the vortex-based diodes. Our mesoscopic geometry hosts both chiral-screening- and vortex-mediated nonreciprocity. We explore their interplay and optimize the SDE.

\section{Results and discussion}
We analyze the SDE in superconducting nanohelices via numerical simulations using the Time-Dependent Ginzburg-Landau (TDGL) formalism\cite{schmid_time_1966} in a finite-element framework\cite{du_global_1994,gao_finite_2023} implemented in COMSOL\cite{noauthor_comsol_nodate, oripov_time-dependent_2020, deenen_periodic_2025}.

The evolution of the order parameter $\psi$ is given by the TDGL equation:
\begin{equation}
    \left( \frac{\partial}{\partial t} + i\kappa\phi \right)\psi = \left( \frac{1}{\kappa}\nabla - i\mathbf{A} \right)^2\psi + (1 - |\psi|^2)\psi,
    \label{eq:tdgl}
\end{equation}
with $\phi$ the scalar potential, $\mathbf{A}$ the vector potential, and  $\kappa=\lambda/\xi$  the Ginzburg-Landau parameter and $\lambda$ and $\xi$ the penetration depth and coherence length, respectively. We solve Eq. \ref{eq:tdgl} together with the Poisson equation for $\phi$ to ensure current conservation:        
\begin{equation}
        \sigma\nabla^2\phi=-\nabla\cdot\left(\frac{i}{2\kappa}\left(\psi^*\nabla\psi-\psi\nabla\psi^*\right)+|\psi|^2\mathbf{A}\right),
    \label{eq:phi}
\end{equation}
where we use dimensionless units (see Supporting Information).

Throughout this work, we will assume the superconductor to be Nb in the dirty limit at fixed temperature $T=0.95T_{\rm c}$, with $T_{\rm c}$ the critical temperature. The penetration depth and coherence length at this temperature are taken as $\lambda = 273$ nm and $\xi = 58$ nm, respectively, giving a Ginzburg-Landau parameter $\kappa=\lambda/\xi=4.7$\cite{rezaev_topological_2020}, and normal conductivity $\sigma=16~\rm{(\mu\Omega~{m})^{-1}}$.

We assume the superconductor is embedded in vacuum, where the superconductor/insulator boundaries apply: $\nabla\psi\cdot\hat{n}=0$, $\mathbf{A}\cdot\hat{n}=0$, $\nabla\phi\cdot\hat{n}=0$.  At the ends of the helices, we inject a transport current $j_{\rm DC}$ and hence apply superconductor/metal boundary conditions: $\psi=0$, $\nabla\phi\cdot\hat{n}=-j_{\rm DC}/\sigma$.  Additionally, we apply the integral constraint\cite{gao_finite_2023}
$\int_\Omega \phi \, d\Omega = 0$, where $\Omega$ denotes the simulated superconductor domain. We neglect screening fields and fields generated by transport currents. Additional information concerning the gauge and the effect of self-fields can be found in the Supporting Information. 

\begin{figure}[h]
\includegraphics{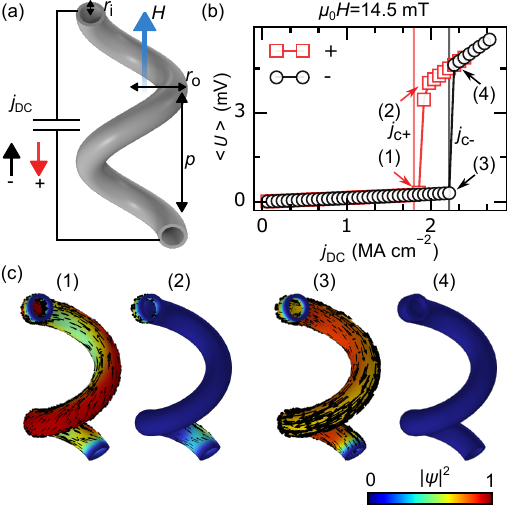}
\caption{(a) Geometry of the superconducting helix and field/current configuration with polarity $\pm$ as indicated. (b) Simulated current-voltage characteristic showing non-reciprocal critical currents. (c) Spatial distribution of $|\psi|^2$ near the transition for both current polarities.}
\label{Fig1}
\end{figure}

Figure \ref{Fig1}a shows a schematic of the main geometry considered in this work. A hollow superconducting helix with minor radius $r_{\rm i}=95.6$ nm, major radius $r_{\rm o}=300$ nm, pitch $p=1.09~\rm \mu m$ and thickness $d=27.3$ nm.  The polarity of the current $j_{\rm DC}$ is indicated with color red (black) for  positive (negative) polarity. To enable the diode effect, we consider an applied field $H$ along the helix axial direction ($z$-direction), consistent with the prediction for magnetochiral anisotropy. The voltage $U$ is computed as the difference in the average scalar potentials found at the top and bottom cross sections $\Gamma_{\rm top}$ and $\Gamma_{\rm bot}$, respectively : $U=\frac{1}{\Gamma,\rm top}\int_{\Gamma,\rm top} dl\,\phi - \frac{1}{\Gamma,\rm bot} \int_{\Gamma,\rm bot} dl\,\phi$ .

We first consider the behavior for low fields, where the transition from the superconducting to normal state is abrupt. Figure \ref{Fig1}b shows the current-voltage characteristic (CVC) of the superconductor under an applied field of $\mu_{\rm 0}H=14.5$ mT. The vertical lines indicate the critical currents $j_{\rm c\pm}$. The critical current is noticeably different for opposite polarities: $j_{\rm c+}\neq j_{\rm c-}$, hence constituting the diode effect. The spatial distribution of the order parameter $|\psi|^2$ close to the transition is shown in Fig.~\ref{Fig1}c. Panels labeled (1) and (2) show $|\psi|^2$ right before and after the transition for positive current polarity, respectively. Before the transition, $|\psi|^2$ is reduced to zero in the inner surface of the helix. In contrast, panels (3) and (4) show $|\psi|^2$ before and after the transition for negative polarity, respectively. Here, $|\psi|^2$ is reduced drastically at the outer surface of the helix.

\begin{figure}[t]
\includegraphics{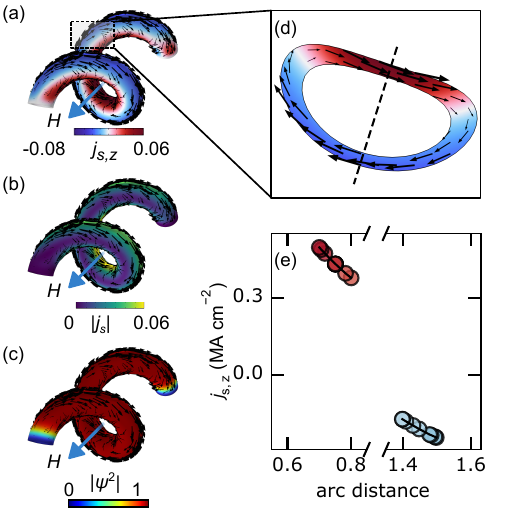}
\caption{The distribution of (a) the $z$ component of the screening current $j_{\mathrm{s,z}}$, (b) the magnitude of the total screening current $|j_{\mathrm{s}}|$, and (c) the magnitude of the order parameter $|\psi|^2$. (d) The distribution of the screening current along the cross-sectional area of the helix (e) $j_{\mathrm{s,z}}$ on opposing sides of the cross-section.}
\label{Fig2}
\end{figure}

To understand the origin of the diode effect, we analyze the spatial distributions of the screening currents $j_{s}$ induced by the applied field $H$. Figure \ref{Fig2}(a-c) shows the $z$-component of the screening current $j_{\rm s,z}$, the magnitude of the total screening current $|j_{\rm s}|$ and the order parameter $|\psi|^2$, respectively. 
The screening current shows an asymmetry between the inner and outer surface. In the inner surface of the helix, screening currents flow along the positive $z$-direction parallel to the applied field $\mu_{\rm 0}H=14.5~\rm mT$. In the outer surface, they flow anti-parallel to the field.  Furthermore, the magnitude of the current density  $|j_{\rm s}|$ is greater in the inner surface than in the outer one. The distribution of $j_{\rm s,z}$ along the cross-section of the helix is depicted in Fig. \ref{Fig2}d, where the arrows are scaled according the $|j_{\rm s}|$.  Figure \ref{Fig2}e
shows $j_{\rm s,z}$ along the dashed line in Fig. \ref{Fig2}d, showing an up/down asymmetry in $j_{\rm s,z}$. 
This asymmetry explains the diode behavior observed in Fig. \ref{Fig1}. The superconducting state becomes unstable once the current density exceeds a critical value $j_{\rm c}$. Under a bias current, the total current at position $\vec{r}$ is the sum of the bias current $\vec{j}_{\rm DC}(\vec{r})$ and the screening current $\vec{j}_{\rm s}(\vec{r})$. When a given bias current is applied parallel to the $z$-direction, the resulting total current density will be higher than when the bias current is applied anti-parallel. This can be seen also in Fig. \ref{Fig1}c, where panels (1) and (3) show a weakening of superconductivity in the inner (outer) surface where $\vec{j}_{\rm DC}$ and $\vec{j}_{\rm s,z}$ are parallel.

\begin{figure*}[h]
\includegraphics{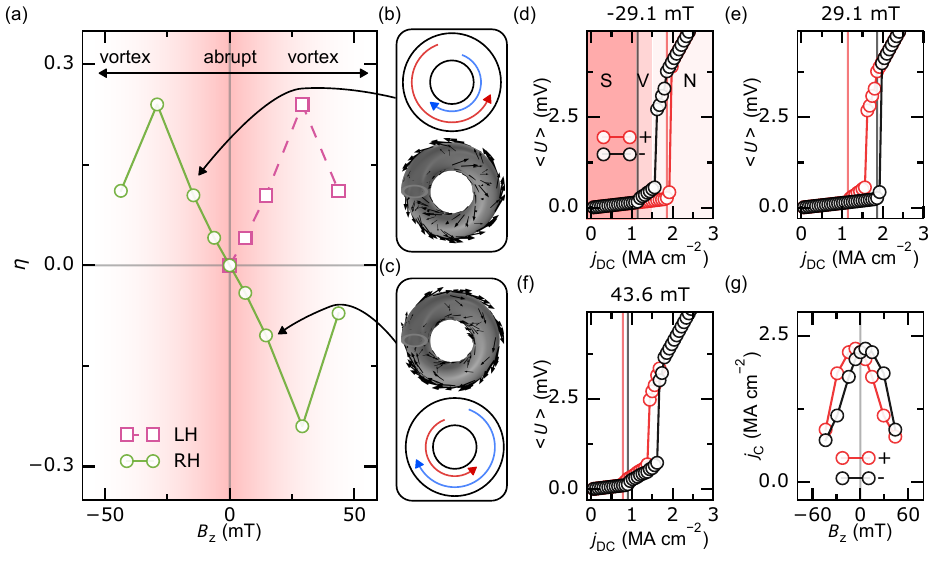}
\caption{(a) The superconducting diode effect efficiency $\eta$ as a function of field for right-handed (RH) and left-handed (LH) helices. The red color gradient indicates a crossover from abrupt SN transition to an intermediate vortex mediated state. (b,c) The simulated screening current distribution at selected fields for RH and LH helices, respectively, as well as a sketch of the screening current loops. The red/blue color coding indicates the sign of $j_{s,z}$, consistent with Figure \ref{Fig2}. (d,e,f) Current-voltage characteristics at selected fields for a RH helix. The coloring in (d) indicates the superconducting state (S), the mixed state (V), and the normal state (N). (g) The critical current density $j_{\rm C}$ as a function of field $B_{\rm z}$ for positive (+) and negative (-) current polarities.}
\label{Fig3}
\end{figure*}

Next, we analyze the field dependence of the superconducting diode effect. To facilitate the analysis, we define the superconducting diode efficiency $\eta=\frac{j_{\rm C+}-j_{\rm C-}}{j_{\rm C+}+j_{\rm C-}}$, where subscripts $+$ and $-$ indicate the polarity of the applied current. Practically, the critical current $j_{\rm C}$ is often defined as the current at which the dissipation (voltage) reaches some threshold value\cite{ruiz_critical_2026}.  Here, we define it as the lowest current at which the derivative $dU/dj_{\rm DC}>0.5~ \mathrm{mv/MA cm^{-2}}$. Thereby, we can evaluate abrupt transitions to the normal states at low fields as well as  the onset of the vortex flow regime for sufficiently high fields.
Figure \ref{Fig3}(a) shows the superconducting diode efficiency $\eta$ as a function of field for nanotubes following either a right handed (RH) or left handed (LH) helix. The sign of the diode efficiency changes as the field is reversed, but its magnitude remains the same. Similarly, changing the chirality from RH to LH changes the sign but not the magnitude of $\eta$. Both observations are consistent with the general expression for MChA. The mechanism behind the sign reversal of $\eta$ is illustrated Fig. \ref{Fig3}b,c. The plot shows a top view of the helix with the supercurrent $j_{\rm s}$ superimposed and a schematic of the corresponding current loops located at either the inner or outer helix surface. The red (blue) color indicates a positive (negative) $j_{\rm s,z}$ component, consistent with Fig. \ref{Fig2}. On the one hand, under reversal of field, both the direction of circulation and the sign of $j_{\rm s,z}$ in the inner-surface current flip, leading to the sign change of $\eta$. On the other hand, when the chirality is changed from RH to LH, the sense of circulation in the azimuthal direction remains the same, but the sign of  $j_{\rm s,z}$ changes.

For small fields, the transition to the normal state is abrupt, and $\eta$ depends approximately linear on the field (Fig. \ref{Fig1}a), consistent with earlier found linear field dependence of the diode efficiency in thin film vortex diodes \cite{suri_non-reciprocity_2022, castellani_superconducting_2025}. At larger fields such as shown in the CVC of Fig. \ref{Fig3}d-f, the transition goes through an intermediate vortex state indicated by (V) in Fig. \ref{Fig3}d. The (V) state is characterized by a slope in the CVC. As the field is increased, first, only one polarity will go through a vortex phase, whereas for the other polarity the transition from S to N is abrupt (Fig. \ref{Fig3}d,e). This maximizes the diode efficiency. At even larger fields, however,  both polarities induce a vortex phase, leading to a smaller diode efficiency. This is seen also when comparing $j_{\rm c+}$ and $j_{\rm c-}$ as a function of field (Fig. \ref{Fig3}g). 

Having discussed the origin and field dependence of the SDE in the helix, we next examine how geometrical parameters influence the diode efficiency. The results are summarized in Figure \ref{Fig4}. Selected CVC and spatial distributions of $|\psi|^2$ at the transition can be found in the Supplementary Information. 
We will focus on the diode efficiency at fields $B_{\rm z}=14.5, 29.1$ mT, which are typically in the abrupt and vortex regime, respectively. The diode efficiency of the reference geometry at these fields are $\eta=10\%,24\%$. Our simulations reveal a pronounced dependence of $\eta$ on the pitch $p$ (Fig.\ref{Fig4}., inset). Smaller pitches, and correspondingly smaller helix angles $\theta=\tan^{-1}\left( p/2\pi r_{\rm o}\right)$, lead to an increased diode efficiency in both the abrupt regime and and the vortex mediated transition. For example at $p=230~\mathrm{nm}$ and $B_{\rm z}=14.5, 29.1$ mT, we find $\eta=15\%, 33\%$, respectively. In contrast, larger pitch lengths tend to decrease the diode efficiency, c.f. $\eta=6\%, 15\%$ at $B_{\rm z}=14.5, 29.1$ mT for $p=1.8~\mathrm{\mu m}$.

Both increasing and decreasing the major radius $r_{\rm o}$ by $0.5\lambda$, i.e., $137$ nm, leads to minor modifications at low fields, but a drop of $\eta$ by a few percent in the vortex mediated regime. Finally, decreasing the minor radius, i.e., the radius of the tube, to $r_{\rm i}=27.3$ nm leads to a lowering of the efficiency to at most 3\%. For such small radii, the system cannot support vortices (whose size $2\xi\approx 116~\mathrm{nm}$), and the transition is always abrupt. Changing the geometry from a hollow tube to a solid wire (spiral) with $r_{\rm i}=27.3~$nm  leads to $\eta=0$. Our study hence shows that artificial chiral superconductors prepared as hollow tubes by additive manufacturing similar to recent 3D magnetic structures \cite{MXu2025} are expected to outperform solid nanohelices as superconducting diodes.
\begin{figure}[h]
\includegraphics{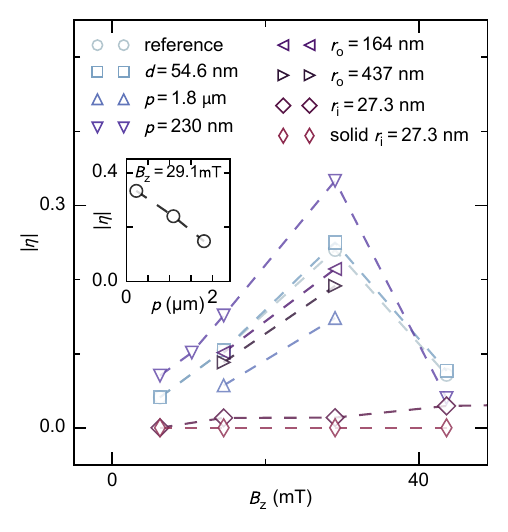}
\caption{Superconducting diode efficiency $|\eta|$ as a function of magnetic field $B_{\mathrm z}$ for various helical geometries. The reference structure has dimensions consistent with Figure \ref{Fig1}a. The labeling indicates deviations from this reference structure. The inset shows the efficiency as a function of pitch $p$.}
\label{Fig4}
\end{figure}

\subsection{Conclusion}

Our results show that the superconducting diode effect can be realized in a mesoscopic chiral geometry, consistent with the magnetochiral anisotropy in noncentrosymmetric superconductors. The origin of the diode effect resides in inequivalent chiral screening current loops at inner and outer surfaces as well as possible vortex nucleation. At low fields, nonreciprocity arises primarily from imbalanced chiral screening currents. At higher fields, asymmetric vortex nucleation further enhances the diode efficiency. For almost all geometries, the diode efficiency $\eta$ is maximized at the field where only one current polarity induces a vortex-dominated transition. The efficiency can be maximized for smaller helix angles. We found a maximum diode efficiency of over 30\%. This efficiency is comparable to recent works which demonstrate the integration of superconducting diodes in cryogenic circuitry:  51\%\cite{ingla-aynes_efficient_2025} and 37\%\cite{castellani_superconducting_2025} under applied fields of 0.5 mT and 4 mT, respectively. 

The nonreciprocal properties described in this work do not depend sensitively on geometrical defects. For low applied fields, the screening current mechanism dominates, which requires only a chiral geometry. In particular, a diode efficiency of over 10\% can still be realized. At higher fields where vortex nucleation plays a role, vortex pinning becomes important, and uniformity of surface properties will be required to achieve reproducible diode efficiency. 

These findings open up new perspectives for integrating top-down fabricated 3D superconducting diodes into multilayer superconducting circuits\cite{makarov_new_2022, fomin_perspective_2022}. Superconducting helices with  sub-100 nm features have already been realized experimentally \cite{cordoba_three-dimensional_2019}. Hollow helices as proposed in this work could be realized using a coated deposition technique such as atomic layer deposition\cite{proslier_invited_2011, femi-oyetoro_plasma-enhanced_2024} on top of a template structure prepared by e.g. FIB, FEBID or two-photon lithography as established recently for a chiral ferromagnetic tube \cite{MXu2025}.

\begin{acknowledgement}

The simulations have been performed using the facilities of the Scientific IT and Application Support Center of EPFL. We acknowledge support by the SNSF via project 10000845.

\end{acknowledgement}

\begin{suppinfo}

Additional details on the numerical procedure, the effect of self-fields, CVCs, and an overview of the extracted critical currents for the geometries discussed in Fig. \ref{Fig4}.

\end{suppinfo}

\bibliography{biblio}

\end{document}